# Impact of Computer-Based Assessments on the Science's Ranks of Secondary Students

Eduardo A. Soto Rodríguez, Ana Fernández Vilas * and Rebeca P. Díaz Redondo

I&C Lab, AtlanTTic Research Centre, University of Vigo, 36310 Vigo, Spain. Correspondence: avilas@det.uvigo.es

**Abstract:** This study reports the impact of examining either with digital or paper-based tests in science subjects taught across the secondary level. With our method, we compare the percentile ranking scores of two cohorts earned in computer- and paper-based teacher-made assessments to find signals of a testing mode effect. It was found that overall, at cohort and gender levels, pupils were rank-ordered equivalently in both testing modes. Furthermore, females and top-achieving pupils were the two subgroups where the differences between modes were smaller. The practical implications of these findings are discussed from the lens of a case study and the doubt about whether regular schools could afford to deliver high-stakes computer-based tests.

**Keywords:** CBAs; PBAs; mode effect; STEM; school scores; teacher assessments

## 1. Introduction

Despite the multiple benefits of computer-based assessments (CBAs) over traditional paper-based assessments (PBAs) stated across the last decades [1–5], the reality is that the adoption of CBAs to deliver high-stakes exams in school settings is still uncommon. The prediction that sooner than later all pupils would have to face the "inexorable and inevitably presence of technologies for assessment [6]" is not confirmed by the results of a recent survey at the European level where 43.7% of secondary students say that they never or almost never use "exercise software, online quizzes and tests in lessons" [7]. Though that situation might be changing since the arrival of the current COVID-19 crisis, we will wonder within this paper to what extent CBAs are actually an affordable technology to regular schools.

The apparent lack of interest in schools for CBAs contrasts with the strong commitment to adopt digital tests in higher education and for administering international large-scale assessments (LSA), as for example, the well-known Programme for International Student Assessment (PISA) and Trends in International Mathematics and Science Study (TIMSS). Nowadays, these on-screen LSA are being administered to school-aged populations who, paradoxically, were barely exposed to that relatively "recent phenomenon" in their schools [8]. This would be anecdotical if it were not for the fact that their outcomes end up carrying direct consequences on the whole educational system in what has been described as the "*new high-stakes phenomenon*" of PISA [9].

Little is known about equivalent deployments to administer high-stakes CBAs in the school setting and, thus, about the impact on school grades. There could be different reasons to explain the apparent lack of interest of school teachers by CBAs. One, more radical, could be that such deployments are feasible for higher education institutions but not for schools where there still are important deficits in technological infrastructures, funds for research and freedom to deploy bottom-up initiatives. Another set of reasons could stem from the fact that the assessment of learning (or achievement) is a critical and sensitive piece in the teaching practice. It has been stated that *"the uncertainties which accompany all change lead educators to maintain the status quo* [10]*"* and, it is also known that a main concern of teachers is *"to know whether they will get the same grade if they adopt a new assessment method" or technology* [11]. These two complementary perspectives, along with the lack of evidence about how CBAs would impact school scores, might explain the reluctance of teachers to undertake strong changes in the administration of high-stakes assessments.

With this paper, we aim to contribute to unpacking that situation from three complementary angles: Firstly, describing as a *case study* how a secondary school deployed complementary high-stakes CBAs to assess students in half of the science subjects taught at the secondary level. This case study would demonstrate that such initiatives are not an exclusive privilege of higher education institutions and of the assessment industry. Secondly, reporting in statistical terms the impact of that initiative on the end-of-course (EOC) grades of two cohorts, sharing and discussing the basics of our methodological approach to compare non-standardized schools scores. Thirdly, supporting the arguments and discussing the findings from the lens of four theoretical frameworks to guide future deployments and further research in this topic.

The section "Assessment and CBAs" includes the theoretical basis and short reviews of literature related to three critical aspects needed to understand the adoption of digital tests in schools: risks and benefits of shifting to CBAs, the outcomes of mode effect studies and the issue of comparability of school scores. The section "The School: A Case Study" describes how a regular school deployed CBAs. The section "Research Questions (RQ) and Methodology" introduces our RQ and the methodology to examine the mode effect with school scores. The section "Results" is divided into two parts: the outcomes of preliminary tests to examine the quality of the raw grades and the answers to our RQ. Our findings, along with the practical implications and the limitations of this work, are extensively commented on in the section "Discussions." Finally, in the section "Conclusions," we summarize this work and share new research directions.

**2. Assessments and CBAs**

In educational contexts, assessment is an ongoing process to collect data about the knowledge, skills, attitudes and beliefs of students to make valid and reliable inferences about how much they learned. Its ultimate goal is "*to reliably and validly distinguish someone who knows something or can do something, from someone who cannot/does not and/or, determine who is a novice, who is an expert, and where someone is on the continuum between the two.* [5]"

Assessments are used across instruction for different purposes. As diagnostic tests, as an "assessment for learning" tool and, as the "assessment of learning" usually required for high-stakes examinations [12]. The stakes (no-, low- and high-stakes) of assessments refer to the degree of consequences for a test-taker of being graded in these tests. In schools, EOC grades are high-stakes to a great extent because they are used for accountability purposes in evaluating performance and determining whether educational targets have been met [9]. They are important to pass a course, to create groups, to apply for admissions and other social and financial benefits such as scholarships, textbooks, transport, free-meal vouchers, etcetera [13].

In this work, the terms "assessments," "tests," "examinations" and "exams" are treated as synonyms, as usually occurs in school settings. CBAs will refer to teacher-made exams delivered through electronic devices (usually computers) and, in our case, automatically marked by the quiz engine of a Moodle platform [14]. On the contrary, PBAs will refer to traditional "paper-and-pencil" exams usually made and marked by the same teacher who taught the subject and administered the exam. The acronyms CBAs and PBAs also indicate the "testing mode" or "test mode" in use. In related literature, equivalent terms include PBT (Paper-based Test), CBT (Computer-based Test), iCBM (Interactive Computer-based Marking), *digital* or *on-screen* assessments are also popular [15–17].

The deployment and use of CBAs described here could be analyzed from the lens of different theoretical frameworks, some extensively documented and others still being developed: For example, the technology acceptance model (TAM) [18] and the technological pedagogical and content knowledge (TPACK) [19] would help to explain the success or failure when adopting an educational technology. Others, more specific but still in development, are the computer-based assessment acceptance model (CBAAM) [20] and the cognitive theory of multimedia assessment (CTMMA) [5], and both can shed light on how to take CBAs to their full potential. Although it is not the scope of this work to dig into these theoretical frameworks, they will be used to back up our arguments and interpret some findings.

*2.1. Advantadges and Risks of Using CBAs*

The quality of tests and of testing instruments should be primarily compared in terms of their degree of validity, reliability, and fairness, particularly when they are used for summative or high-stakes purposes. From the view of the TAM model, to understand the benefits and risks of adopting CBAs, it is important to check how well digital exams deal with these factors in comparison with traditional PBAs.

Validity is the "sine qua non" of assessment. Without evidence of validity, assessments in education have little or no intrinsic meaning [21]. Among the notions of validity, "Construct validity" is the most important as it is an essential measure of the perceived overall validity of a test. Cronbach and Meehl (1955) defined construct as "some postulated attribute of people, assumed to be reflected in test performance (p. 283) [22]". In simpler words, "construct validity" broadly refers to whether a test measures what it claims to measure. An alleged advantage of CBAs is that they would allow measuring *new* constructs (e.g., solving complex scientific problems interacting with a simulator in real-time), which do not fit well or are not possible with traditional PBAs [23–25]. For example, the chance to insert dynamic stimuli, such as multimedia files or interactive objects into digital items, would tend to increase the level of "construct representations [26]". Therefore, if we assume that "The best construct is the one around which we can build the greatest number of inferences (p. 288)[22]", CBAs would offer more advantages and capacity than PBAs to draw advanced inferences. In short, to assess better. To illustrate this argument, it is worth mentioning the effort invested within PISA and TIMSS to develop in each cycle innovative interactive instruments to measure abilities which "*have never been tested before*" [27,28].

Besides of valid, tests must also be reliable and fair [9]. Test fairness basically refers to provide equal opportunities to all test-takers [29], and test reliability refers to the degree to which an assessment tool produces stable and consistent results under similar circumstances and, thus, yield accurate test scores [30]. It is known that human scoring, besides being labor intensive and time-consuming, is also prone to validity and reliability problems [31]. Therefore, another double advantage of using automatically marked CBAs would be that they are free of common human-marking biases [32,33] and that machine's scores are more accurate than the teacher's scores [34].

Most of the risks and threats of shifting to CBAs stem from the old concern that "computers affect testing" [35] and the subsequent risk of "*measuring other skills different from those intended, as for example the student's skills to use technological devices*" [36]. For more than a decade, the lack of familiarity with electronic devices was the main threat reported in literature to question the validity and fairness of CBAs. However, a very recent work did not find evidence that computer access or familiarity was nowadays more important for taking digital- than for taking paper-based tests [37]. Current discussions about the validity of CBAs are more centered on other issues, as for example, the impact of unappropriated designs of the multimedia and interactive elements embedded in digital items [5] or about how to ensure secure administrations of CBAs. These risks might lead to false positives (i.e., when CBAs help test-takers more than expected) or to false negatives (i.e., when multimedia contents induce unexpected "noise" causing wrong answers) outcomes. On another hand, a remarkable advantage of CBAs is the capacity to make a systematic capture of log data during the assessment experience. Conveniently analyzed, these data could provide new valuable insights. For example, to shed light on behavioral aspects about how test-takers complete a task [38], to better track the student's progress, to improve the teaching practice over time and to push toward more effective integration of learning and assessment [39]. In sum, it has been highlighted here some remarkable advantages of CBAs over PBAs supported by the opinions of other experts in this field [4,40] but also some risks or threads that should be taken into account by new adopters before accepting this technology.

*2.2. Testing Mode and Achievement*

The effect of the testing mode ("mode effect") in achievement has been broadly investigated with data from standardized LSA [41,42]. In the absence of equivalent studies in school settings, the outcomes of LSA studies are the best reference in literature to examine whether the testing mode affected the measurement of achievement in school populations and in the specific field of science. LSA consists of a massive administration of tests under standardized conditions where large samples of test-takers sit a state, national or international exam the same day, covering identical or equivalent content but either in paper or on-screen. Then, researchers compare the outcomes to observe whether the scores earned under each mode are different. For these studies, the design and selection of items, the construction of tests and all the logistics and rules required to administer these tests are led by external assessments experts, while school teachers are simply observers without opinion or influence in these events. An important flaw found in LSA as PISA and TIMSS is that the lack of pupil's interest or motivation to perform well along with their "academic endurance" to sustain the level of effort until the end of the test can give unreliable results [43,44].

It would be worth having references from similar analyses supported by data from equivalent deployments in school settings. Unfortunately, a search for related literature in the Education Resource Information Center (ERIC) database and Google Scholar using, among others, the terms "CBA" (or any of its other equivalent terms), "implementation" (or deployment) and "secondary school" (or secondary education) did not bring useful references to be cited here. The few references found are not useful for comparisons for distinct reasons: Most are pilot studies that lasted for a limited period of time instead of for at least the full duration of a course. In others, the focus of the assessments was not science, or the educational level where it was implemented was not secondary, or CBAs were used exclusively for formative instead of summative purposes. In sum, nothing was found in literature about long-term "mode effect" studies under non-standardized conditions, with school grades as the main outcome variable and implemented with the existing resources of average schools.

Therefore, we will review the outcomes of eight LSA "mode effect" studies published since 2005 and selected because they were administered to populations of secondary students and concerned with the educational area of science, which is the scope of our study. Two of these studies are meta-analyses of prior works, and six correspond to research studies based on more recent data from LSA.

In four of these selected studies, the authors reported that the CBA and PBA scores were either equivalent or one mode slightly easier than the other but without reach statistical significance [45–48]. In three, the authors found that the scores in PBAs had been higher than in CBAs, but only in one case, the difference was reported as statistically significant [49]. In the other two studies that correspond to two well-known international LSAs: a field trial of PISA 2015 conducted in Germany, Sweden and Ireland [50] and a study to examine potential mode effects in TIMSS [51]; the authors reported similar findings. However, in a meta-analytic review of fourteen prior studies, it was found that in ten, both tests were comparable overall; in three, the science tests were harder in PBAs mode; in one, the CBAs version was more difficult. They concluded that, in general, science tests were comparable across the two modes [52].

Some of these studies also reported other interesting findings. For example, Kingston (2009), in a meta-analytic review of 81 studies between 1997 and 2007, concluded that, while the grade (or educational level) had no effect on comparability, disciplines, on the contrary, did have an effect. However, the effect sizes for science disciplines were very small [53]. On another hand, Kim and Huynh (2007) found that the evidence of comparability between PBAs and CBAs in Biology tests existed at the item-, subtest- and whole test-level [47]. These two studies are important to back up our research because they suggest that neither the grade nor the science subjects are expected to affect the comparability of scores in both modes. Although recent studies suggest that the magnitude of mode effects differ much more across students [51] and alert administrators to expect large differences in "experience with *digital assessment*" [54] and to find students "*at risk*" *of performing differently on digital and paper tests* [28], we think that these differences are more likely due to a lack of exposure and training of some pupils with CBAs.

Another issue that deserves further research relates to the interaction between the participants' ability levels and performance across test modes. While some studies report that ability is not a predictor of performance differences [55], others found opposite results. For example, Clariana and Wallace (2002) found that individuals classified as having high content attainment or higher prior academic attainment, especially in math and science, were advantaged by computerized tests [56]. However, another study suggested that CBAs might affect low-achieving students more positively by significantly increasing their motivation and performance [57]. Similarly, it was also found that in some particular cases, females and males might perform differently depending on the testing mode [35,40,50,58]. A more extensive review of the literature concerning the effect of the testing mode by gender and the outcomes of a by-gender analysis with our data has been published in a prior paper [59].

In sum, none of these cited works are concluding references to our research for four reasons: First, they are based on administering identical standardized versions of the same test (double-testing), and that approach rarely occurs in schools. Second, the raw scores in our research are not single test scores but the EOC grades awarded to pupils across their secondary education. Third, in our study, we aim to compare the effects of mode on ranks instead of on raw scores, so we will convert the EOC grades into percentile rank scores (PRSs). Fourth, the entire implementation, this is, the development of items, the construction of the test and the administration of these assessments were not led by assessment experts but by secondary teachers across school years. Nevertheless, all that body of research from LSA is important because it shows the "mode effect" from a laboratory perspective. To conclude, what most of the aforementioned studies show is that, overall, the average "mode effect" found was either null or small. Our data will allow investigating all these topics at distinct levels, this is, at cohort, gender and ability (achievement) levels.

2.3. Are School Grades Comparable?

School grades are the most important single metric to report pupil's achievement annually. To what extent school grades are comparable is another controversial issue that affects this work because we will take the EOC grades of pupils as the outcome variable to investigate the mode effect. According to some literature, school scores are neither valid nor reliable because they tend to be the result of teacher's merge judgements of academic and non-academic factors [13,60], which are prone to "grade inflation" [61–63] and affected by well-known human biases. It is a fact that some teachers are extremely strict, while others are extremely lenient in awarding marks. Moreover, either consciously or unconsciously, they are under influences operating at the marking stage, as for example, the "halo effect," the "order effect" or other personal beliefs or criteria, which ultimately make it extremely difficult to compare their grades [32,33]. Furthermore, the scores of teacher-made assessments, which are the norm in schools, are considerably less reliable than of the standardized tests administered in LSA.

However, other authors claim just the opposite. This is that the school scores collected and awarded by teachers in the course of their teaching are important to "integrate the many elements of performance behaviors required in dealing with authentic assessment tasks" [12] and, thus, to increase the validity of student assessment. Proponents of this opposite view call for more research within assessment boards to achieve "a common currency" to interpret school grades from a "holistic view" [64,65].

Another controversial issue relates to whether it is rational to compare school subjects. Those who hold a simplistic view of school subjects as "apples and pears," that is, as inherently different, there would not be room for comparisons. However, other experts in the field of educational measurements share opposite arguments to justify such comparisons. From a broadminded vision of comparability, all the school subjects measure (at least to some extent) "general aptitudes" and thus, their outcomes could be legitimately compared [66]. In a special issue about comparability published by Cambridge Assessment, it is argued that the scores of different school disciplines can be compared if there is evidence that "they share common features and use" [67] or "on the basis of a particular attribute" [68].

The issue of comparability is important in our research because we aim to compare school grades earned in different subjects. To overcome the limitations of such comparisons will narrow the scope of the study to the educational level of secondary, to the field domain of science and to within-subject analysis. That is, we will assume that the six science disciplines taught across secondary (see Table 1) share a common "linking construct" called in PISA "scientific literacy [69]", are "sufficiently unidimensional" to be compared and, share a similar rank of difficulty for students of secondary education [66,70]. Moreover, research carried out in the UK in 2008 suggested that at A-level, the science subjects "physics, chemistry and biology were generally well aligned to one another in performance [71]" and exhibit the highest marking consistency among all the school subjects [72].

**Table 1.** A list of science subjects by course and school year taught to each cohort across secondary education, including the delivery mode of exams: CBAs or PBAs.

| ID Code | Science Subject | Grade (Course) | School Year | Cohort | Listed As | Test Mode |
|---|---|---|---|---|---|---|
| NS1_14 | Natural Science (I) | Grade 7 (1st) | 2013–2014 | Y10 | MANDATORY | PBA |
| NS2_15 | Natural Science (II) | Grade 8 (2nd) | 2014–2015 | Y10 | MANDATORY | PBA |
| BG3_16 | Biology and Geology | Grade 9 (3rd) | 2015–2016 | Y10 | MANDATORY | CBA |
| PC3_16 | Physics and Chemistry | Grade 9 (3rd) | 2015–2016 | Y10 | MANDATORY | CBA |
| BG4_17 | Biology and Geology | Grade 10 (4th) | 2016–2017 | Y10 | OPTIONAL | CBA |
| PC4_17 | Physics and Chemistry | Grade 10 (4th) | 2016–2017 | Y10 | OPTIONAL | PBA |
| NS1_15 | Natural Science (I) | Grade 7 (1st) | 2014–2015 | Y09 | MANDATORY | PBA |
| NS2_16 | Natural Science (II) | Grade 8 (2nd) | 2015–2016 | Y09 | MANDATORY | PBA |
| BG3_17 | Biology and Geology | Grade 9 (3rd) | 2016–2017 | Y09 | MANDATORY | CBA |
| PC3_17 | Physics and Chemistry | Grade 9 (3rd) | 2016–2017 | Y09 | MANDATORY | CBA |

Furthermore, to limit the effects of the aforementioned biases, the raw EOC grades will be converted into PRSs and compared exclusively at the within-pupil (paired-tests) level. This decision is justified by three reasons: The first is based on the guidelines to interpret computer-based tests published in 1986 by the American Psychological Association, which can be summarized as:

*"scores may be considered equivalent, or comparable, when the rank orders of individuals closely approximate one another and when the score distributions are approximately the same, or have been made approximately the same through rescaling* [73]"

The second is that, in the same document, the author, a reputed researcher in this field, invites to not consider only mean score differences but also "differences in rank ordering" [73]. The third is based on the expectation that pupil's ranks will remain stable from year to year unless substantial changes occur [67] or, put in other words, "*if a cohort seems not to have changed much from a previous cohort, the grade distribution should not differ much either*" [74].

In sum, we see in this introduction that all aspects of assessments are susceptible to be improved using appropriate technologies. Besides, a variety of assessment methods is desirable [40], and CBAs, with their advantages and risks, can be a feasible, complementary and useful tool to improve the assessment practices in schools and to report achievement in science subjects at the level of secondary education. In LSA and under standardized administrations, the outcomes of achievement tests in sciences are not different when they are sat as CBAs or PBAs. However, little is known about what would be expected after similar long-term implementations in regular schools. To attract new practitioners, it is important to look at the mode effect on school grades, but this is still a controversial issue that needs further research. To move ahead, it is important to find case studies, start collecting real-life data systematically and reporting analysis to better understand this *new* phenomenon.

## 3. The School: A Case Study

The raw EOC grades for this study were provided by the department of sciences of a Spanish urban school, supported by public funds in their levels of mandatory education, which offers educational services to pupils who live nearby as established by the regional administration. Both the school and the surrounding areas are, on average, representative of many other schools and neighborhoods in the country. The main reason to select this school was that they had completed a technological deployment to administer high-stakes CBAs with their own resources in three of six science disciplines taught in secondary education.

Prior research called to look at the "role of school characteristics" to better understand "how will technology-rich school environments impact student performance on computer-based testing [47]" and with a special focus on cases where regular teachers are at the center of these innovations:

*"it is vital that teachers become active agents for change, not just in implementing technological innovations, but in designing them too"* [10].

That is the case in this school, where, years before awarding these grades with CBAs, some science teachers began to deliver occasional no- and low-stakes digital quizzes covering distinct topics taught in science lessons. After years of practice, they became more and more familiar with the technology for CBAs and aware of the risks, advantages and opportunities to transform their assessment practices from the belief that "assessment *is* learning" [75]. They also understood that prior practice and familiarity with the digital testing tool were imperative before moving ahead to deploy CBAs for high stakes purposes [17,76,77]. Thanks to the Quiz Module of the Moodle platform installed at the school, which includes algorithms to mark answers automatically and to deliver scores and feedback, some teachers began to create item banks and to assemble and administer digital tests. Subsequent improvements to adapt school infrastructures to host secure CBAs and the expertise gained with regular low-stakes tests allow them to progress into what is called the *second-staged* of a CBAs integration characterized by:

*"incremental changes, including innovative item formats, the automation of various assessment processes and early attempts to improve the measurement of constructs (or aspects of constructs) that have proven difficult to measure using paper-and-pencil tests."* [73].

At the end of the 2015–2016 school year and after three years of successive pilot phases to start adopting high-stakes CBAs examinations in distinct disciplines, the school began to administer only CBAs in three science subjects (50%) while the other three subjects remained administering only PBAs. It means that, since that year and along the whole course, the "CBAs disciplines" administered exclusively digital tests and the PBAs disciplines only paper-based tests and both made by teachers.

### 3.1. The School Assesment Policy

Another detail, which is key in this research, is that the department of sciences had established rules to award grades in science subjects limiting the weights of cognitive and non-cognitive outcomes. They stated that 80% of the end-of-term (EOT) and EOC grades had to come from invigilated exams covering the knowledge and skills established by the national standards for each discipline. Other additional records usually gathered by teachers (i.e., group interaction, behavior, effort, participation, etc.) could never account for more than the remaining 20% of these final scores. That rule applied equally to subjects administering CBAs or PBAs. The issue is that, in school settings, these non-cognitive outcomes are usually assessed through direct observation, and the scores are given based on teachers' qualitative judgements, which are not always weighted equally. Though, in a recent work, the authors claimed that these "*Teacher assessments of achievement are as reliable and stable as test scores at every stage of the educational experience*" [78]. In short, the important aspect to highlight here is that this mandatory grading scheme with a fixed weight for exams provides a stable framework to fulfil the comparisons of school scores intended in this work. Otherwise, without that common rule, we would never know to what extent these EOC scores are comparable.

### 3.2. The Data Set of EOC Grades

This study is based on a data set of anonymized raw EOC grades from two cohorts named Year 9 (Y09) and Year 10 (Y10), which also included the gender. These cohorts were chosen for two reasons: First, they had been already exposed to both testing modes across their secondary studies. Second, they were not affected by any of the two educational reforms undertaken in Spain around that period. These two circumstances brought a unique opportunity to study the mode effect in a real setting because both cohorts followed the same curriculum, and they already had grades awarded in subjects examining in both modes.

Table 1 shows the list of the six science disciplines included in the Spanish curriculum of secondary at that time. Four are mandatory and taught in grades 7, 8 and 9, and two are offered as optional subjects in grade 10th. The first two letters of ID codes correspond to the subject's name, the following number indicates the course where it was taught, and the two digits after the underscore refer to the school year where they were taught. For example, NS1_14 refers to a mandatory subject named Natural Sciences of 1st course of Secondary (Grade 7), which used PBAs to examined pupils of cohort Y10 in the 2013–2014 school year.

A pupil of Y10 who had completed the four years of secondary at the school and was enrolled in all (mandatory and optional) science subjects would be examined with PBAs in Y7 and in Y8, with CBAs in Y9 (two mandatory subjects) and, with one subject in PBAs and another in CBAs in Y10. In Spain, the EOC grades in these disciplines are recorded in the student's gradebooks as ordinal numbers (without decimals) and ranging from 0 (rarely used) to 10. The next two tables show the descriptive stats of the raw EOC grades earned in each discipline by 118 students of cohort Y09 (Table 2) and 100 of cohort Y10 (Table 3), both enrolled at the school at the end of the 2016–2017 school year.

**Table 2.** Descriptive stats of EOC raw grades earned by the students of cohort Year 9 (Y09) in the four mandatory science subjects studied until that course.

| Cohort Year 9th | N | Mean | SD | Median | SE |
|---|---|---|---|---|---|
| BG3_17 (CBA) | 118 | 6.10 | 2.42 | 6.19 | 0.23 |
| PC3_17 (CBA) | 118 | 6.01 | 2.59 | 6.02 | 0.24 |
| NS2_16 (PBA) | 110 | 5.70 | 2.88 | 5.27 | 0.27 |
| NS1_15 (PBA) | 105 | 5.93 | 2.69 | 6.38 | 0.26 |
| BG3_17 (CBA) | 118 | 6.10 | 2.42 | 6.19 | 0.23 |

**Table 3.** Descriptive stats of EOC raw grades earned in mandatory and optional (*) science subjects studied by the students of cohort Year 10 (Y10).

| Cohort Year 10th | N | Mean | SD | Median | SE |
|---|---|---|---|---|---|
| * BG4_17 (CBA) | 38 | 6.02 | 2.91 | 6.05 | 0.47 |
| * PC4_17 (PBA) | 70 | 5.88 | 2.69 | 5.56 | 0.32 |
| BG3_16 (CBA) | 88 | 5.96 | 2.78 | 5.84 | 0.30 |
| PC3_16 (CBA) | 88 | 6.00 | 2.81 | 5.51 | 0.30 |
| NS2_15 (PBA) | 86 | 5.77 | 2.89 | 6.67 | 0.31 |
| NS1_14 (PBA) | 83 | 6.09 | 2.65 | 7.18 | 0.29 |

**4. Research Questions and Methodology**

The equivalence between CBAs and PBAs scores has been a central concern in education as CBAs increases [79]. The impact of individual differences on the testing experience, and so the statistical equivalence of scores, still needs to be considered, and that will be the goal of our research questions.

As noted, to our knowledge, there are no prior studies comparing the effects of shifting from CBAs and PBAs modes in school grades earned in science. That gap in literature leaves us orphans of useful references to our methodology. The huge differences between the long-term (continuous) assessments of a course reflected in the school grades and the one-time standardized tests administered in LSA require a novelty methodological approach to draw valid conclusions.

In this work, we aim to know if the PRSs received by pupils within their respective cohorts in science subjects examining with CBAs were equivalent to those obtained in subjects administering PBAs. It is important to remark that the comparisons will be at a rank level and not at a raw grade level. Moreover, these comparisons will be done at a within-student level; that is, we will run paired-tests on the PRS of each pupil awarded under CBAs and PBAs mode. If both testing modes rank students similarly or with differences falling within a predefined margin of equivalence, we will report a lack of statistical impact; otherwise, we will report the magnitude of the observable impact.

*4.1. Research Questions*

**RQ1:** *Is there an overall statistical equivalence between the EOC ranking scores obtained in science disciplines solely delivering CBAs and those solely delivering PBAs?*

**RQ2:** *Is there statistical equivalence of these scores at the cohort level?*

**RQ3:** *Is there statistical equivalence of these scores at the gender level?*

**RQ4:** *Is there statistical equivalence of these scores at the achievement level?*

*4.2. Methodoloy*

To this study, we needed the EOC grades of pupils with at least one grade earned in a PBAs subject and another in a CBAs subject, and both earned at the school. After excluding 16 school newcomers (eight from Y09 and eight from Y10) who did not meet the requirements, we ended up with three main samples or groups: one of 110 students of Y09 (52 females and 58 males), another of 92 students of Y10 (40 females and 52 males) and combining both cohorts into a new group called ALL (N = 202). Then, we created four additional subgroups dividing the pupils of the group ALL by quartiles. It means that those at or below the 25th percentile (first quartile) populated the group of low-achievers (LA), those within the second quartile were labelled as medium-achievers (MA), those within the third quartile as high-achievers (HA) and, finally, those above the 75th percentile were labelled as top-achievers (TA). The first three *groups* (Y09, Y10 and ALL) were used to investigate the impact of the mode effect at overall, within-cohort and gender levels. Additionally, the last four subgroups (LA, MA, HA and TA) at the achievement level. The specific details of the by-gender analysis are available in a published paper presented in the international congress [59]. Our statistical study comprises four steps described in Figure 1. First, we ran a set of preliminary analyses of the raw EOC grades in science disciplines of each cohort (Y09 and Y10) to estimate their degree of "convergent validity," internal consistency and to what extent they reflected a unidimensional latent trait. Then, we checked whether the raw EOC scores of different disciplines were correlated within cohorts as would be expected because they belong to a well-defined field domain, as it is science. Secondly, we ran tests of internal consistency to see how reliable these raw EOC grades were at measuring the achievement in science. Internal consistency is usually measured with Cronbach's alpha, a statistic calculated from the pairwise correlations between the items of a test. However, in our context, these items will be the distinct science disciplines studied by each cohort. Coefficients of internal consistency of at least 0.90 are expected for professionally developed high-stakes standardized tests; however, for classroom exams, it is desirable to reach a reliability coefficient of 0.70 or higher [80].

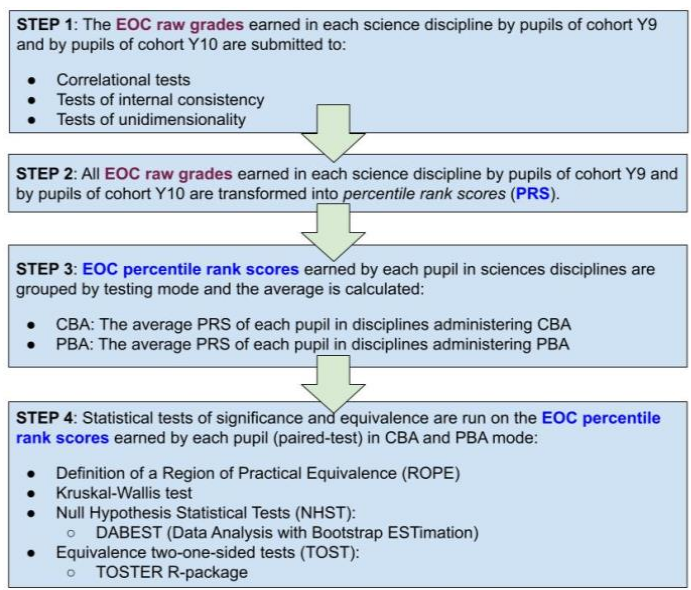

**Figure 1.** A diagram of the four steps of our methodological approach.

Although it was not strictly required for our analyses, we ran an additional statistical test using the "dim" function within the *R*-package PSYCH [81] to confirm that the set of raw EOC grades earned by each cohort reflected a unidimensional latent trait as we had assumed they were measuring. In sum, the goal of these three preliminary tests was to prove that our data set of raw EOC grades were valid and reliable measurements of student's achievement in science, and therefore, they could be used for subsequent analysis.

To find signals of a mode effect, we decided to compare PRSs instead of raw scores. Thus, the second stage consisted of transforming the raw EOC grades into PRSs. To carry out these transformations, we calculated the PRS of each pupil in each discipline respect to their cohorts. That is, pupils were ranked in each discipline according to their rank position with respect to the rank position of the rest of classmates within the same cohort.

In the third step, we aggregated the PRSs in science by the testing mode to get a unique CBA and a unique PBA score for each pupil. These calculated CBAs and PBAs scores resulted from averaging the PRSs achieved in disciplines administering CBAs and in disciplines administering PBAs. These CBA and PBA scores of pupils will be the main outcome variables of our research.

The fourth step consists of running a set of statistical tests on the EOC PRSs to answer the research questions. If any significant change is found in the rank order of pupils will be attributed to the testing mode [67,74]. The statistical tests selected for this step will say whether the paired differences between the CBA and PBA scores at the within-subject level are significantly different from zero and, in such case, how large (magnitude).

The first test will be a test of correlation to see how strong is the association between the PRSs earned by each pupil in both modes. The second test will be a Kruskal–Wallis test to get an overview of the main effects of each variable (COHORT, GENDER and MODE). Finally, we will run null hypothesis statistical tests (NHST) followed by tests of equivalence to find out if a mode effect exists on pupil's ranks. To qualify a difference between modes as large enough to be reported as equivalent, we need to define what would be a reasonable limit to that equivalence. In other research fields, it is used a so-called region of practical equivalence (ROPE) which can also be found with other names as for example: "small effect size of interest" (SESOI), "indifference zone," "margin of non-inferiority" or "range of equivalence," among others [82]. Unfortunately, in this field, we could not find similar works from where to get a reference for a ROPE for school scores. Therefore, we decided to establish an interval range of ±5% for our ROPE because that is the default value in many other research fields. This means that, in this work, when the mean of the differences between CBAs and PBAs scores fell within an interval of ±5%, it will be reported as statistically equivalent [82,83].

To estimate how large are the differences between testing modes, we selected a statistical tool called DABEST (Data Analysis with Bootstrap ESTimation) available for R, Python and Mathlab. This software provides new graphical representations of several effect sizes to compare observations from the same individuals, including illustrative plots to make estimation graphics accessible to all scientists [84]. An additional reason to use this software was that school EOC grades do not always meet the assumptions of normality and homogeneity required by classic parametric tests. According to the authors, when using the bootstrap resampling of DABEST, "there is no need to assume that our observations, or the underlying populations, were normally distributed because, thanks to the Central Limit Theorem, the resampling distribution of the effect size will approach normality" [85]. Besides, estimation statistics move beyond *p*-values and focus on the magnitude of the effect and its precision. We will look at the 95% confidence interval (CI) of the paired mean differences reported by DABEST. If both CIs ends fall within the ROPE limits, then we could not reject the equivalence. Nevertheless, we will double-check the outcomes of DABEST against the outcomes of TOSTER, a specific R-package to test equivalence [86]. TOSTER uses a specific procedure called two one-sided tests (TOST), which allows testing for equivalence between modes and rejects the presence of an effect smaller than the ROPE limits [83]. Both tests should provide complementary insights; therefore, if any dissonance is found between them, it will be reported. Otherwise, it should be assumed that they pointed in the same direction.

## 5. Results

This section presents the results as follows: Firstly, the outcomes of our preliminary tests (correlation and internal consistency) ran on the raw EOC grades (V.1). Secondly, the results of the statistical tests ran on the PRSs of the different groups (V.2 and V.3) are described in step 4.

*5.1. Preliminary Analysis*

Tables 4–6 show the descriptive stats of these CBA and PBA percentile rank scores variables for each group of study.

**Table 4.** Descriptive PRSs statistics of group ALL and cohorts Y09 and Y10 in CBA and PBA mode.

|  | CBAs | | | PBAs | | |
| --- | --- | --- | --- | --- | --- | --- |
|  | Y09 | Y10 | ALL | Y09 | Y10 | ALL |
| Pupils | 110 | 92 | 202 | 110 | 92 | 202 |
| Mean | 5.968 | 5.878 | 5.927 | 5.749 | 5.755 | 5.751 |
| Std. Error of Mean | 0.231 | 0.263 | 0.173 | 0.256 | 0.256 | 0.181 |
| Std. Deviation | 2.421 | 2.522 | 2.461 | 2.680 | 2.459 | 2.575 |
| 25th percentile | 3.898 | 4.157 | 3.898 | 3.684 | 3.895 | 3.895 |
| 50th percentile | 6.186 | 5.674 | 5.674 | 5.827 | 5.808 | 5.827 |
| 75th percentile | 7.818 | 8.211 | 7.839 | 7.712 | 7.695 | 7.712 |

**Table 5.** Descriptive PRSs statistics of Female and Male pupils in CBA and PBA mode.

|  | CBAs | | PBAs | |
|---|---|---|---|---|
|  | Female | Male | Female | Male |
| Pupils | 92 | 110 | 92 | 110 |
| Mean | 6.385 | 5.545 | 6.184 | 5.390 |
| Std. Error of Mean | 0.273 | 0.216 | 0.287 | 0.226 |
| Std. Deviation | 2.616 | 2.266 | 2.751 | 2.371 |
| 25th percentile | 4.093 | 3.898 | 4.414 | 3.847 |
| 50th percentile | 6.992 | 5.212 | 6.325 | 5.200 |
| 75th percentile | 8.452 | 7.161 | 8.473 | 6.966 |

**Table 6.** Descriptive PRSs statistics of subgroups of achievers in CBA and PBA mode.

|  | Low-Achievers | | Medium-Achievers | | High-Achievers | | Top-Achievers | |
|---|---|---|---|---|---|---|---|---|
|  | CBAs | PBAs | CBAs | PBAs | CBAs | PBAs | CBAs | PBAs |
| Pupils | 51 | 51 | 51 | 51 | 50 | 50 | 50 | 50 |
| Mean | 0.695 | 1.006 | 2.264 | 2.726 | 4.849 | 4.687 | 7.816 | 7.839 |
| Std. Error of Mean | 0.067 | 0.092 | 0.133 | 0.145 | 0.156 | 0.209 | 0.161 | 0.173 |
| Std. Deviation | 0.476 | 0.654 | 0.950 | 1.033 | 1.105 | 1.478 | 1.142 | 1.226 |
| 25th percentile | 0.434 | 0.533 | 1.989 | 2.006 | 4.205 | 3.810 | 7.222 | 6.840 |
| 50th percentile | 0.641 | 0.908 | 2.094 | 2.799 | 5.089 | 4.680 | 7.906 | 7.784 |
| 75th percentile | 0.661 | 1.327 | 2.756 | 3.625 | 5.341 | 5.882 | 8.609 | 8.865 |

All Spearman and Pearson's pairwise tests of correlation among disciplines indicated that the raw EOC grades within cohorts Y9 and Y10 were highly and positively correlated at $p < 0.01$. These results were interpreted as evidence of convergent validity, hence assuming that they were measuring a common construct, namely the achievement in "scientific literacy" [69]. The tests of internal consistency on the same grades yielded high Cronbach's alpha values in Y9 ($\alpha$ = 0.94, 95% CI (0.92, 0.95)) and in Y10 ($\alpha$ = 0.95, 95% CI (0.94, 0.97)). These results were clearly above the reliability coefficient of 0.70 and declared "desirable" for classroom exams [80]. Therefore, we assumed that the EOC grades passed the required test of reliability. Finally, using the *R*-package PSYCH [81], we could confirm that these grades reflected a single unidimensional latent trait as predicted. In sum, these preliminary analyses suggested that the EOC raw grades were valid and reliable measurements of pupil's achievement in science at the secondary level. Hence, they were transformed into PRSs (second step) and finally aggregated by testing mode (third step). Their descriptive statistics by cohort, gender and achievement are displayed in Tables 4–6.

*5.2. Overall Mode Effect on PRSs*

At the overall level (group ALL), it was found that the PRSs earned in disciplines delivering CBAs and PBAs were highly correlated (Pearson's *r* = 0.84 and Spearman's *rho* = 0.83) and that the CBAs scores explained the 71% of the variance in PBAs scores. Similar results were found at the cohort level (Figure 2).

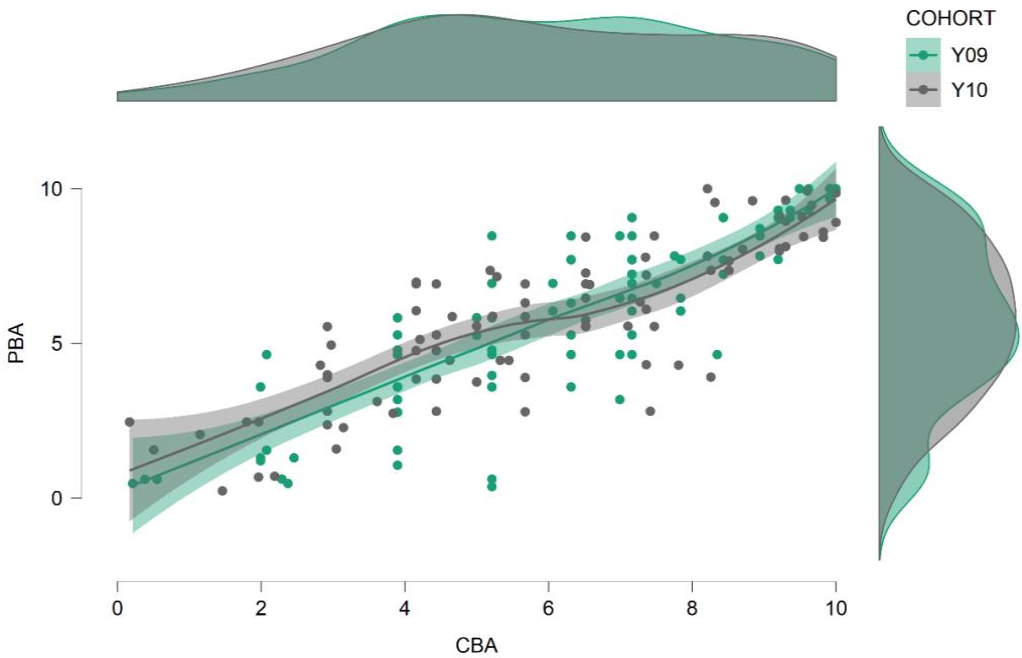

**Figure 2.** A scatterplot of PRSs of 202 pupils (group ALL) by MODE (PBA and CBA) and by COHORT (Y09 and Y10).

A Kruskal–Wallis test ran on the variables of group ALL ($n$ = 202) only reported a main statistical effect of GENDER on the pupil's PRSs (chi-squared = 13.05, df = 1, $p$-value < 0.001). This means that the PRSs of females are significantly higher than the PRSs of males. The other factors (MODE and COHORT) had not a statistical impact (Table 7).

**Table 7.** Kruskal–Wallis Test.

| Factor | Statistic | df | $p$ |
| --- | --- | --- | --- |
| MODE | 0.350 | 1 | 0.554 |
| COHORT | 0.055 | 1 | 0.815 |
| GENDER | 13.049 | 1 | <0.001 |

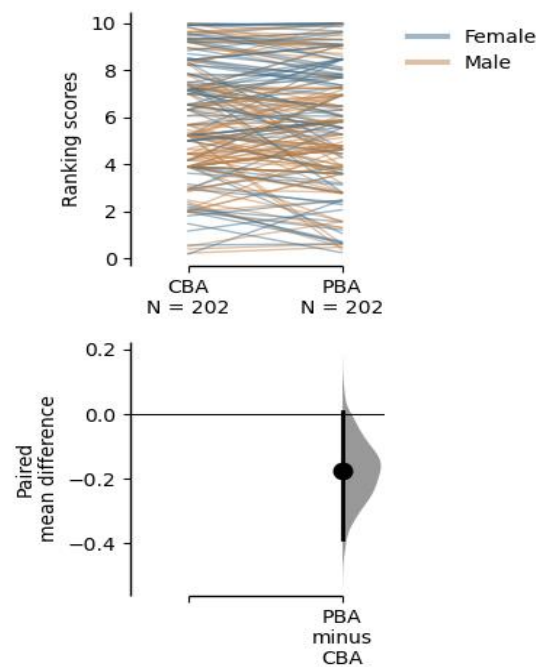

**Figure 3.** Gardner–Altman estimation plot for the paired mean difference between PBA and CBA scores of 202 pupils (group ALL).

In RQ1, we aimed to investigate whether, at the overall level, the PRSs of each pupil in CBAs and PBAs testing modes differed from a statistical point of view. Our results show that the paired mean difference of PRSs between disciplines examining in PBAs and CBAs mode was −0.176 (95%CI −0.386, 0.009). That difference is not statistically different from zero ($p > 0.05$), and the 95% CI falls completely inside the ROPE (see Figure 3).

To illustrate the paired mean difference between PBAs and CBAs scores, the DABEST package shows Gardner–Altman estimation plots as in Figure 3. In these plots, both groups are plotted on the left axes as a slopegraph: each paired set of observations is connected by a line. The paired mean difference is plotted on a floating axe below as a bootstrap sampling distribution. The mean difference (−0.176) is depicted as a black dot; the 95% CI is indicated by the ends of the vertical error bar. The TOSTER package also confirmed that the observed effect is statistically not different from zero and statistically equivalent to zero ($p = 0.000739$).

*5.3. Mode Effect within Groups*

At the within-cohort level (RQ2), it was found that the PRSs in both modes were not statistically different from zero (see Figure 4). In Y09 ($n = 110$), DABEST reported a PBAs minus CBAs difference of 0.297(95%CI 0.00508, 0.587) with a $p$-value of the two-sided permutation t-test equal to 0.0508. In cohort Y10 ($n = 92$), the difference was −0.00168 (95%CI −0.303, 0.306) with a $p$-value of the two-sided permutation $t$-test equal to 0.99.

As it is observed in Figure 4, the top end of the vertical error bar in cohort Y09 exceeds the limits of our ROPE, suggesting a likely lack of equivalence. However, the outcome of the TOST procedure in these data concluded that the observed effect was statistically not different from zero $t(109) = 1.650$, $p = 0.102$ and statistically equivalent to zero ($p = 0.0189$). In short, both testing modes awarded equivalent PRSs.

At the gender level (RQ3), we found similar results in both sexes (Figure 5). The paired mean differences between the PBAs and CBAs scores of 92 females is 0.0835 (95%CI −0.225, 0.397), and the $p$-value of the two-sided permutation t-test is 0.604. That difference in the group of 110 males is 0.225 (95%CI −0.0708, 0.489), and the $p$-value of the two-sided permutation $t$-test is 0.116.

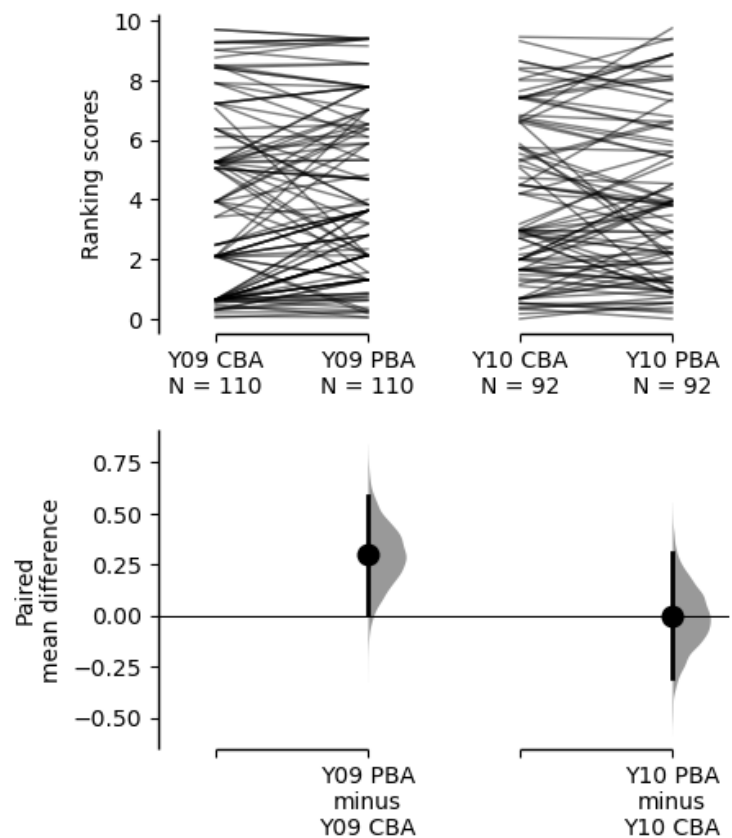

**Figure 4.** Gardner–Altman estimation plot for the paired mean difference between PBA and CBA scores of 110 pupils (cohort Y09) and 92 pupils (cohort Y10).

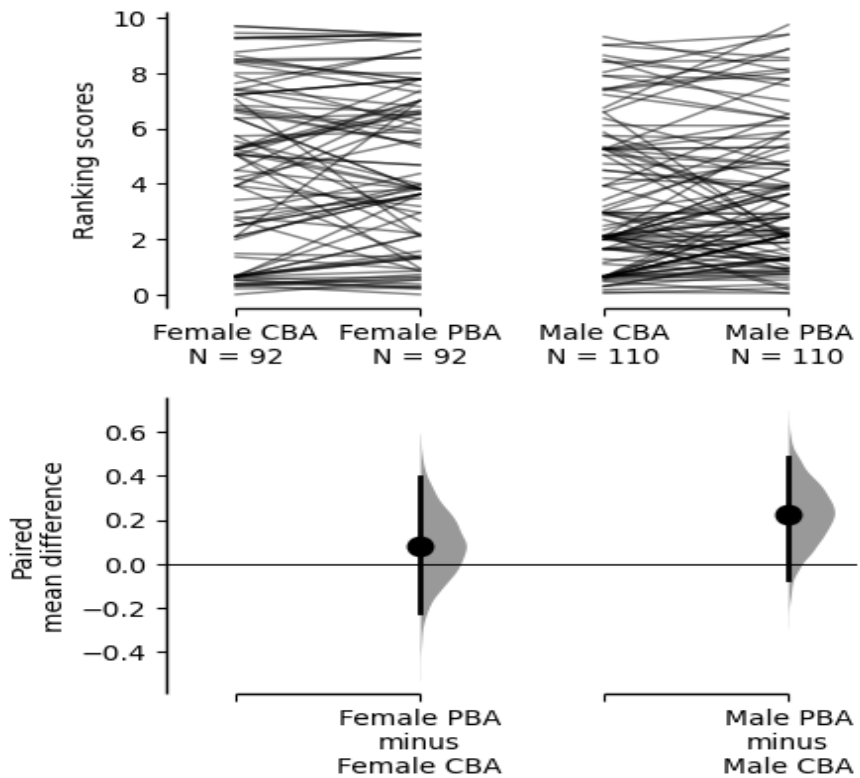

**Figure 5.** Gardner–Altman estimation plot for the paired mean difference between PBA and CBA rank scores in science subjects of 92 females and of 110 males.

Both results indicate that the mode differences are not statistically different from zero, and both 95% CI fall inside the ROPE. The TOST outcomes also conclude that the observed effects at the gender level are statistically not different from zero and statistically equivalent to zero.

Finally, at the achievement level (RQ4), results were mixed (Figure 6). On one side, the paired mean difference of PRSs earned by HA (−0.162 (95%CI −0.732, 0.436)) and by TA (0.0228 (95%CI −0.273, 0.368)) in science subjects administering CBAs and PBAs did not differ from zero. On another side, the same difference was significant in the subgroups of MA (0.462 (95%CI −0.0217, 0.86)) and LA (0.311 (95%CI 0.0835, 0.516)). A small discrepant result about significance was found in LA where the DABEST result was significant, but the NHST reported by TOSTER did not ($p = 0.073$).

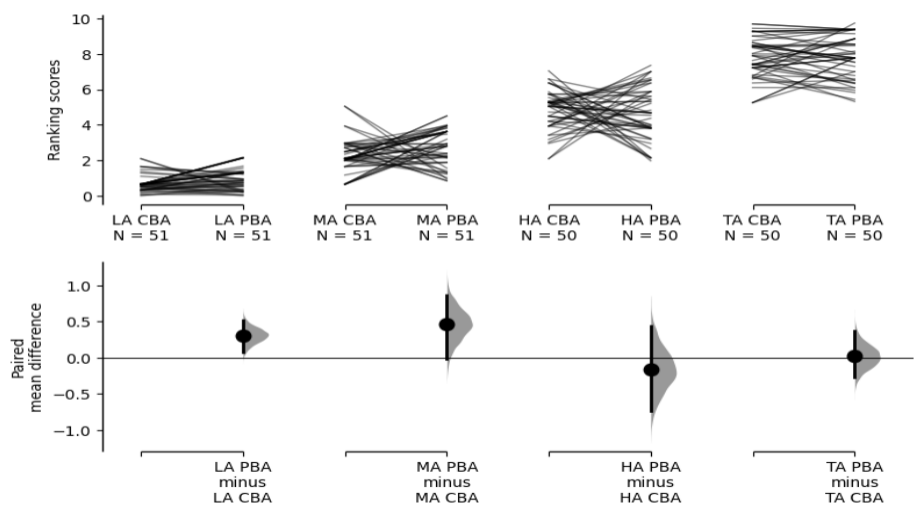

**Figure 6.** Gardner–Altman estimation plot for the paired mean difference between PBAs and CBAs rank scores in science subjects of 51 low-achievers (LA), 51 medium-achievers (MA), 50 high-achievers (HA) and 50 top-achievers (TA).

The subgroup of TA is the only one where the magnitude of the difference between modes felt totally inside the ROPE, while the other three subgroups exceeded the limits. The TOST confirmed that the difference between testing modes was statistically not equivalent to zero in LA ($p$ = 0.324) and in HA ($p$ = 0.214), but statistically equivalent in TA ($p$ = 0.0015) and, surprisingly, in MA ($p$ = 0.046) despite the length of the 95%CI.

Fortunately, our sample sizes to answer the RQ1, RQ2 and RQ3 were large enough to detect with enough statistical power (>90%) the small size effect of interest (SESOI), that is, the ±5% defined as our ROPE. However, unfortunately, the sample sizes of LA, MA and HA did not permit achieving the desirable 80% level of statistical power, and thus, the results of these groups should be cautiously interpreted. However, this did not occur with the subgroup of 50 TA, where we did get a statistical power over 80% to detect the SESOI. The importance of the result found in the subgroup of TA will be extensively commented on in the next section.

*5.4. Mode Effect at Within-Subject Level*

The last analysis was made within cohort Y10 to look at the mode effect at the within-subject level. This was possible because the EOC grades of Y10 included scores earned by 66 students in two Physics and Chemistry courses (PC3_16 and PC4_17) and under distinct testing modes. Moreover, another subgroup of 38 students was enrolled in BG4_17 and PC4_17 at the same time and thus, exposed to CBAs and PBAs testing modes, respectively. The paired mean difference between the PRSs earned in PC4_17_PBAs and PC3_16_CBAs was 0.181 (95.0%CI −0.31, 0.613), which is not statistically significant ($p$ = 0.454). Similarly, between BG4_17 and PC4_17, the difference was −0.305 (95.0%CI −0.713, 0.194) and the $p$-value= 0.161. These two analyses at the within-discipline level yielded results consistent with our main outcomes. However, further research is needed to confirm this finding for two reasons: One is that the statistical power for the sample size of 38 pupils to detect the size effect of interest in this work was too low (57.58%). Another is that comparisons among mandatory and optional subjects should be done with caution to control for potential effects of selection biases.

**6. Discussion**

Are CBAs affordable technology to administer high-stakes examinations in regular schools or just for higher education institutions and the assessment industry? The answer to that question is not trivial because the current evidence shows that CBAs are not yet in the toolboxes of teachers if we look at what EU pupils say [7] and to the lack of references in literature. If we admit that it is affordable, why is the adoption so low? In this paper, we commented from the lens of theoretical frameworks many factors that can explain that situation. From the lack of the required infrastructures to the perception of little or minimal utility of this tool by any of the actors in the education system (educational authorities, school leaders, parents, teachers or students). The case study presented in this work describes an uncommon but real-world deployment of CBAs in secondary education, which suggests that similar initiatives can be replicated in other schools.

Besides introducing the challenges (advantages and risks) of adopting CBAs and describing the school as a case study, we also investigated the impact of their implementation on schools scores. Once EOT or EOC final grades are being awarded through CBAs, it is important to look for signals of a mode effect. Otherwise, it could be claimed that CBAs are either not valid, reliable or fair, and thus, being rejected forever. To know whether a mode effect exists we have to compare these grades, but that is not possible unless they had been awarded under an exclusive testing mode and under a common assessment policy, and all these prerequisites are unusual in school settings. However, the deployment of CBAs at the school of our study and the marking policy they followed did permit us to run such comparisons using the EOC grades in science disciplines as the dependent variable.

A preliminary analysis revealed that these EOC raw grades of 202 pupils, no matter the testing mode used to award them, were highly correlated and consistent, which is in line with what other works had also found in science grades [71,72]. To carry out this research, we followed a particular methodological approach (see Figure 1) to minimize some issues concerning the use of school scores. For example, we decided to compare PRSs instead of raw grades in an attempt to avoid some marking biases and teacher influences acknowledged in literature that might threaten the analysis [13,32,33,60,63]. Furthermore, our research is narrowed to the educational field of science, where there is plenty of evidence of high consistency in grades compared to any other curricular subjects [72] and four years of mandatory secondary education. Besides, all comparisons were paired comparisons; that is, we did not compare at the between- but at the within-subject level. This means that our inferences about the mode effect at overall, cohort, gender or achievement level are based on computing firstly single individual effects.

To report the impacts in statistical terms, we followed a perspective based on estimation statistics more focused on the magnitude of the effect and its precision, instead of relying solely on a single *p*-value [84]. Without references in literature, we decided to adopt a ROPE of ±5% to judge whether the PRSs earned in CBAs and PBAs disciplines were actually equivalent. From the perspective of practitioners (school's teachers and leaders), having evidence of equivalence is much more informative than a simple *p*-value of non-statistical difference. In this sense, it would be important that future works justify what would be a reasonable ROPE to compare school scores.

The EOC school grades are a sort of "composite index" where teachers reflect the overall performance across a whole course, and therefore, they are of the highest stakes. These records are annually saved in the student's gradebooks and are good proxies to track the progress of pupils across basic education. Any substantive change in the distribution of these grades at the group level should be investigated to detect anomalies, and that is what we did here comparing ranks. That investigation is key to ensure that the assessment tool works as expected and without strong dissonances between the outcomes of new and traditional assessment practices. That is why it is argued that "the burden of proof is on the CBAs side to demonstrate that it produces equivalent scores [76]" in both modes. That would be the price to pay initially to boost the confidence and the acceptance among the school stakeholders and to limit the distress on those who decide to use it. As commented, this is another take-away from the case study, which fits within the guidelines of the TAM to foster a successful acceptance of technology [18].

With our research questions, we aimed to know whether the mode effect existed for particular pupils, first, at the overall and cohort levels and then at the gender and achievement levels. The call of Escueta et al. (2017) to investigate "to what extent particular populations are affected for delivered modes [87]" means that this topic is not closed in LSA and much less at the school level. Overall, the results of RQ1, RQ2 and RQ3 suggest that the decision of the science department of examining solely with CBAs in three out of six science subjects taught across secondary education had not a meaningful impact on the PRSs of students. It was found that the rank position of pupils had not been altered by the assessment mode, no matter whether the analyses are made at overall or cohort levels. These findings are consistent with other studies previously cited and based on datasets from standardized assessments [45–48,50–52]. Moreover, we did not get evidence of CBAs harming the performance of female students more and favoring male students, as has been suggested in some studies [49,88]. Rather, we found just the opposite: the PRSs of boys were a bit higher in PBA mode, whereas girls performed similarly in both modes [59]. Our outcome matches the PISA results on science reported by Jerrim et al. (2018) about the lower scoring differences found in girls with respect to boys. On another hand, to undertake the study at the achievement level, we had to divide pupils into four equivalent groups (LA, MA, HA and TA) according to their positions in ranks. The resulting sub-samples contained between 50 to 51 students, and only in the subgroup of TA, there was enough statistical power to detect the size effect (±5%) stated in our ROPE. It means that while our outcomes at gender and at TA levels are supported by enough statistical power, the outcomes at LA, MA and HA do not, so they must be interpreted with caution.

In school contexts, apart from the obvious differences in attainment, TA and LA also differ in the homogeneity of profiles. Within LA, we expect a variety of pupils from those with small, medium and strong special needs to others who are demotivated or do not want to study. However, the TA group includes those who usually share an interest in learning, tend to cover the whole material before an exam, are more resilient, have more grit and tend to perform higher in most subjects [89]. Another aspect to take into account is that the grades within the subgroup of LA are usually less reliable than in any other subgroup because they are not always awarded under the same criteria. For example, it is acknowledged within teacher's circles that, in some cases, they avoid giving too low scores to not discourage pupils and keep them engaged to try hard in subsequent attempts; therefore, we cannot know to what extent the scores of LA are reliable. With regard to the results of MA and HA, we see large variability likely because they are usually less consistent in their efforts and, thus, in their performances across courses and within particular subjects. In sum, the lack of enough statistical power alongside the mentioned features of LA, MA and HA makes it difficult to interpret their results; therefore, we could not reject or confirm that CBAs more positively affect low-achieving students as other authors had stated (Nikou and Economides, 2016).

However, two remarkable findings of this study are precisely the results of females (Figure 5) and TA (Figure 6) groups where we can observe that their 95%CI are inside the ROPE, well centered around zero and also have the shortest distances between the extremes (error vertical bars) of all groups. That similar distribution can be easily explained by the fact that females are generally top-achievers, so they tend to be of the subgroup TA as well. We argue that this result is important to defend that the CBAs administered by the school were reliable because these (females and TA) are the groups where a lack of equivalence would be much more difficult to justify. If we are right, this finding would send a positive message to teachers thinking to adopt CBAs.

Overall, our research questions suggest that the transition followed by the school to examine with alternative testing modes had no impact on how students were ranked. The computed within-subjects differences were always inside the limit stated as of practical equivalence (±5%). We think that this finding is important because it sets up the first reference for further works aiming to study real-life transitions between testing modes in the school setting and to compare paper- and digital-based assessment instruments. Our dataset included neither the tests nor the items used in the science assessments. Further research should also look at other important related aspects of CBAs. For example, at how well particular disciplines (biology, physics, chemistry, geology) within a common domain (e.g., science) and of distinct educational levels fit CBAs. This would allow us to see if the absence of differences at the subject level reported in the literature can be confirmed with school scores [47,53]. Another research direction could be to study the *mode effect* at topic-level, to see whether some content fits better or worse with CBAs. In parallel, other important analyses should be done at the whole-test and at the item level where the Item Response Theory (IRT) can shed light on the quality of these items and tests and provide new robust basics for comparisons.

Moreover, our results should not be interpreted without considering the singularities of the deployment of CBAs carried out by the school. The case study shows that it is key to follow some rules in the implementation, which are also solidly backed up in literature. For example, about the leading role of teachers to carry out a successful transition or about that the transition to high-stakes CBAs should be done after a long-term exposition to frequent low-stakes tests. To the school staff, the key to success lies much more in the proactive behavior of teachers to gain expertise and confidence working with the testing tool to develop their own items, digital forms, and deliver frequent low-stakes tests both at school and as online homework than in having infrastructures. In their view, the pathway to accept CBAs for high-stakes examinations will be easier if teachers believe in the benefits of frequent testing, also known as *retrieval practice effect* [90], and in "assessment *is* learning [75]". A final step for these teachers was to learn to make the arrangements to assure secure administrations of digital exams in the computer rooms of the school.

It turns out that all these basic empirical ideas are consistent and well-aligned with the postulates of the TPACK and TAM theoretical frameworks and with how other researchers described the behavior of teachers when facing critical challenges [10,11,18,19]. From different perspectives, these works emphasize the "central role of teachers" to promote the acceptance of new technologies in educational settings. It means that, in addition to mastering some basic technological skills, it is teachers who must primarily perceive the usefulness of the tool and need to have enough time to practice and learn with it to end up transforming their assessment practices.

Another singularity of the schools refers to the familiarity of students with CBAs. The team of teachers who led that transition assured that their pupils were fully accustomed to taking digital tests before they started to administer high-stakes CBAs. This means that they were familiar with the testing tool, the types of items, the navigation modes and devices used for these summative CBAs. That is precisely the path recommended in the literature to successfully adopt CBAs for high-stakes examinations [17,28,37,54,76,77]. It is also aligned with the guidelines to achieve a successful acceptance among practitioners [18,20]. Another key detail is that these teachers had learnt through years of practice how to design their own digital items and how to assemble testing forms with them. Furthermore, they knew how to make the arrangements to assure secure administrations of digital exams in the computer rooms of the school. This reveals that the final shift to high-stakes CBAs was made only when they already had the resources, perceived the usefulness of the tool and mastered a *technological pedagogical knowledge,* as it is described within the TPACK model [19]. From their view, some CBAs implementations could fail because they might be based on a top-down model of innovation where defenders mistakenly think that an investment in infrastructures will be automatically followed by a rapid adoption. Rather, they believe the opposite. That is, to succeed, resources should grow parallel to the freedom to introduce small changes and to the availability of time to develop skills practicing with low-stakes assessments.

To close this section, we refresh the initial question: Are CBAs an affordable technology to administer high-stakes examinations in regular schools? As we see it, it is not clear what will end up happening with the use of high-stakes CBAs in schools. If it is perceived as a big challenge or as an unaffordable task for schools, it will likely end up being hired or leased to external providers as other school services nowadays. However, if the educational authorities see it as a valuable tool for regular teachers, more resources should be given to deploy infrastructures and to offer updated professional training. That training would need to be rethought and perhaps inspired by a new cognitive theory of multimedia assessments (CTMMA) [4,5] to help practitioners to bring the tool up to its full potential.

## 7. Conclusions

To the best of our knowledge, this is the first work where the implementation of high-stakes CBAs in a secondary school is presented as a case study. We also studied the impact of the testing mode in the EOC grades of 202 pupils earned in science subjects across four years of secondary education for the first time. To undertake this work, we had to raise and discuss some issues that should be addressed before using schools scores as the main outcome variable.

The lack of published papers about the effect of the testing mode on schools contrasts with its abundance in higher education, where CBAs are regularly used and are the focus of advanced research. That gap could suggest that the requirements administer regular high-stakes CBAs exceed the current capacity of schools. However, our case study would rather demonstrate the opposite. That is that average schools can implement CBAs with their own resources and from a bottom-up innovation model.

Despite the classical limitations of all ex post facto (observational) studies, this work intends to be a first and holistic reference to this complex phenomenon where many research branches converge. It is notorious that this topic deserves more research both at a theoretical and an empirical level to boost confidence and acceptance among school stakeholders.